\documentclass[12pt]{article}
\usepackage{amsfonts,graphics,amsmath,amssymb}
\usepackage{lscape}
\usepackage{epsf}
\newcommand{\half}{{\textstyle{\frac{1}{2}}}}
\newcommand{\thalf}{{\textstyle{\frac{3}{2}}}}
\newcommand{\ZZ}{{\mathbb Z}}
\newcommand{\lra}{\longrightarrow}
\newcommand{\llra}{\leftrightarrow}

 \newcommand{\be}{\begin{equation}}
\newcommand{\ee}{\end{equation}}
\newcommand{\bea}{\begin{eqnarray}}
\newcommand{\eea}{\end{eqnarray}}

\newcommand{\JN}{J^{N}}
\newcommand{\jn}{\Delta{J^{N}}}
\newcommand{\ji}{\Delta{J^{i}}}
\newcommand{\jk}{\Delta{J^{k}}}
\newcommand{\jim}{\Delta{J^{i-1}}}
\newcommand{\jkm}{\Delta{J^{k-1}}}
\newcommand{\jdue}{\Delta{J^{2}}}
\newcommand{\ai}{A_{i}}
\newcommand{\aik}{A_{i,k}}
\newcommand{\aikd}{A_{i,k}^{\dagger}}
\newcommand{\aid}{A_{i}^{\dagger}}
\newcommand{\jp}{J_{+}}
\newcommand{\jm}{J_{-}}
\newcommand{\s}[3]{\sum_{#1=#2}^{#3}}

\newcommand{\jph}{J_{+,H}}
\newcommand{\jmh}{J_{-,H}}

\newcommand{\jpv}{J_{+,V}}
\newcommand{\jmv}{J_{-,V}}
\newcommand{\aikm}{A_{i,k;m}}
\newcommand{\aikmd}{A_{i,k;m}^{\dagger}}
\newcommand{\aim}{A_{i;m}}
\newcommand{\aimd}{A_{i;m}^{\dagger}}
\newcommand{\jpm}{J_{+,m}}
\newcommand{\jmm}{J_{-,m}}
\newcommand{\upa}{\uparrow}
\newcommand{\doa}{\downarrow}

\begin{document}

\pagestyle{empty}
\setcounter{page}{0}

\vfill

\begin{center}
\vspace{25mm} 

{\Large \textbf{Mutation model for nucleotide sequences \\  based on crystal basis}}

 \vspace{18mm}

{\large C. Minichini ,   A. Sciarrino$^{\dag}$} \\
\emph{Dipartimento di Scienze Fisiche, Universit{\`a} di Napoli
``Federico II'' \\
and  $^{\dag}$I.N.F.N., Sezione di Napoli \\ %
Complesso Universitario di Monte S. Angelo\\ %
Via Cintia, I-80126 Naples, Italy}

\end{center}

\vspace{20mm}

\begin{abstract}
 A nucleotides sequence is identified, in the two (four)  letters alphabet, by the the labels of a  vector state of an  irreducible representation of $\mathcal{U}_{q \to 0}(sl(2))$
 ($\mathcal{U}_{q \to 0}(sl(2) \oplus sl(2))$). A 
 master equation for the distribuion function is written, where the 
intensity of 
 the one-spin flip is assumed to depend from the variation of the 
labels of the state. In the two letters approximation, the 
   numerically computed equilibrium distribution for short sequences is  nicely fitted by a Yule
   distribution, 
which is the observed distribution
 of the ranked short oligonucleotides frequency in DNA. The four letter alphabet description, applied to the codons, is able to reproduce the form of the fitted rank ordered usage frequencies distribution.
 \end{abstract}

 
 \vspace{2cm}

  
\begin{flushright}
preprint DSF 14/2005
\end{flushright} 

\vfill  
 
\noindent
email:  \texttt{minichini@na.infn.it} $\;\;\;$ \texttt{sciarrino@na.infn.it} \\
Corresponding author: \texttt{A. Sciarrino - Tel +39-081676807 - Fax +39-081676346}

 \newpage
\phantom{blankpage}
\newpage
\pagestyle{plain}
\setcounter{page}{1}
\pagestyle{plain}
\setcounter{page}{1}
\baselineskip=17pt

\section{Introduction}

Mutations  in DNA play a very important role in the theory of evolution. 
DNA and RNA are build up as sequences of  four basis 
or nucleotides which are usually identified by the letters:  C, G, T, A (T being replaced by
 U in RNA), C and U  (G and A) belonging to the purine family, denoted by R (respectively to the pyrimidine family, denoted by Y).
 Therefore in the case of genome sequences each point in the 
 sequence should be identified by an element of a four letter alphabet 
 or by a set of two binary values.  In a simplified treatment one identifies each 
 element according to the  purine or pyrimidine nature, reducing so to a  two letter alphabet or  to a
 binary set. 
Genetic mutations, i.e. modifications of the DNA genomic sequences, play
a fundamental role in the evolution.  They include changes
of one or more than one nucleotide, insertions and 
deletions of nucleotides, frame-shifts and inversions. In the present 
paper we consider only the point mutations, for a review see  (Li Wen-Hsiung, 1997). These are usually modeled by 
stationary, homogeneous Markov process, which assume: 1) the nucleotide 
positions are stochastically independent one from another, which is clearly 
not true in functional sequences; 2) the mutation is not
depending on the site and constant in time, which  ignores the existence 
of ``hot spots" for mutations  as well as the probable existence of 
evolutionary spurts ; 3) the nucleotide frequencies  
are equilibrium frequencies.   
In the simplest model  one can think of,  all the mutations are assumed reversible and with equal rate,    therefore only one parameter rules all the transitions. This is clearly a very rough approximation and indeed more complicated models have been proposed  depending on more parameters. The most  general one, with not reversible transitions, depending on the type of the nucleotide undergoing a mutation and on the kind of mutation,  requires 12 parameters. However all these models are based on the assumption that the transitions are  not depending of the neighbour nucleotides.
 In the early  nineties was realized that the intensity of  point mutations is  really depending on the context where they happen (Blake R., Hess and Nicholson-Tuell, 1992; Hess, Blake J. and Blake R., 1994) and
in the last decades an increasing amount  of data  in genetic research has  provided further evidence that there is indeed a not negligible effect of the nearest neighbors as well as an effect of the the whole sequence, see e.g.  (Arndt,  Burge and Hwa, 2002).
In the more simplified descriptions,  where the elements of the two chemical families (purine and pyrimidine), the four nucleotides belong to,  are identified, a correspondence is made between the nucleotides and the elements of a binary  set. It follows that  the mutations are  mathematically modelised as transitions between  binary labels sequences.  As a binary alphabet is equivalent to
spin variables, it is clear that the spin  approaches,  extensively studied in physics, have a natural application in the  theory   of molecular biological evolution. Indeed since 1986, when
 Leuth\"{a}usser (Leuth\"{a}usser, 1986, 1987) put a correspondence between the Eigen model
  of evolution  (Eigen 1971;  Eigen, McCaskill and Schuster, 1989)  and a two-dimensional Ising model, many 
  articles have been written representing biological systems as  
  spin models.
In  (Baake  E., Baake  M. and Wagner,  1997) it  has been shown that the parallel mutation-selection model can be put in 
correspondence with the hamiltonian of an Ising quantum chain and in
(Saakian and Hu, 2004) the Eigen model   of evolution has been 
mapped into the hamiltonian of one-dimensional quantum spin chains.
In this approach the genetic sequence is specified by a sequence of 
spin values $ \pm 1$.   In more refined models the correspondence is made between the four nucleotides and a set of two binary labels, see (Hermisson, Wagner and Baake M., 2001) for a four-state quantum chain approach.
 The main aim of the works using this approach, see (Baake  E., Baake  M. and Wagner,  1998), 
(Wagner, Baake  E. and Gerisch,  1998), (Baake E. and Wagner, 2001), (Hermisson,  Redner and Wagner, 2002),  is to find, in different landscapes, the mean ``fitness" and the 
 ``biological surplus", in the framework of biological population 
evolution.  As standard  assumption,   the strength 
of the mutation is assumed to depend from the distance between two 
sequences, which is identified with the Hamming distance. We recall that the
Hamming distance between  two strings of binary labels is given by the number of sites with different 
labels. Moreover usually it is assumed that the mutation matrix 
elements are vanishing for Hamming distances larger than 1, i.e. for 
more than one nucleotide changes. 
The Hamming distance assumption is clearly
unrealistic in the domain of genetic mutations, so the only justification for its use is  
 that this assumption generally allows to solve the problem exactly in the one point mutation 
scheme or to find more tractable numerical solutions. 
 For example the 
mutation between the sequence $\ldots GUGU-ACAC \ldots$ and the  
sequences, both differing of one unit, in the Hamming distance, from the 
original one,
$\ldots GUUU-AAAC \ldots$ and  $\ldots GGGU-ACCC \ldots$ implies respectively a 
 change in the free energy, at standard conditions, of $ \approx -0.89$ 
 kcal/mol and of $ \approx +0.8$ kcal/mol, see (SantaLucia, 1998).
 To assume  these transitions  equally probable is clearly a rough approximation.
 Let us note that we use the term transition in a physical general sense. 
 In biology \textit{transition} is a mutation from a purine (pyrimidine)
 to a  purine (pyrimidine), \textit{transversion} is a mutation from a purine
 (pyrimidine) to the other family.
 So,  in the above specified simplified assumption, the 
 transitions have to be really understood as biological \textit{transversions}.
At our knowledge there has been no attempt to apply spin models to 
obtain the observed equilibrium distribution of oligonucleotides 
in DNA. Martindale and Konopka  (Martindale and Konopka, 1996)  have,  indeed, remarked that 
the ranked short (ranging from 3 to 10 
nucleotides) oligonucleotide frequencies, in both coding and 
non-coding region of DNA, follow a Yule 
distribution. We recall that a Yule distribution (Yule, 1924)  is  given by
\be
 f = a \, {n^k}\, {b^n}  \label{eq:Yule}
 \ee
  where $n$ is the rank and $a$, $ k < 0$ and $b$ are 3 real parameters.
 In order  to face this problem, in this paper we propose a  
 spin model  where effects of neighbours (not only the nearest  ones) and on the
whole sequence context is taken into account.  
  To this aim, we  build up a  quantum and classical spin model in which the strength 
 of the transition matrix does not depend only from the number of different 
 symbols (Hamming distance) between two sequences, but in some sense
  also  from the position of the changed symbols and from the whole distribution
  of the nucleotides in the sequence. 
   In this paper, we assume that the transition matrix 
 does not vanish only for total spin flip equal $\pm 1$, induced by the 
 action of a single step operator, which generally  is equivalent to one 
 nucleotide change.
Let us recall some phenomenological aspects of mutations. 
 From observations on the characters of spontaneous mutations,
  it seems possible to point out
some common  features of almost every studied process.
 These can
be resumed in the following points:

\begin{itemize}

\item the mutation rate of a nucleotide depends on nature of its
      first neighbouring ones;

\item mutations occur more frequently in purine/pyrimidine alternating
      tracts;

\item \textit{transitions} are more frequent than \textit{transversions};

\item mutations mainly interest dinucleotides \textsc{CG}.

\end{itemize}

In modeling mutation mechanism, only paying attemption on the
difference between purines and pyrimidines (so that we only consider
\textit{transversions}), we take only into account the first  two of the four points above listed. 
 
 In a context slightly different,    it has been 
  remarked (Frappat, Minichini, Sciarrino and Sorba, 2003) that the rank of codon usage probabilities follows a universal law, that is independent of the biological species, the  rank ordered distribution $f(n)$ 
 being nicely fitted by a sum of an exponential part 
and a linear part.  Of course 
the same codon occupies in general two different positions in the rank 
distribution function for two different species, but the shape of the function is the same.  
 More specifically, for each biological species,  codons are ordered following the decreasing order of the values of their
usage probabilities, i.e.  the codon with rank $n =  1$ corresponds to the one with highest 
value of the codon usage frequency,
codon  with rank $n = 2$ is the one corresponding to the next highest value of the codon usage frequency, and so on.    
 In  that article $f(n)$ was plotted  versus the
rank and  was well fitted by the  following function
 \begin{equation}
f(n) = \widehat{\alpha} \, e^{-\widehat{\eta} n} \, - \, \widehat{\beta} \, n \, + \, \widehat{\gamma} \;,
\label{eq:bf}
\end{equation}
where $0.0187 \leq \widehat{\alpha} \leq 0.0570$, $0.050 \leq \widehat{\eta} \leq 0.136$, 
$0.82
\; 10^{-4} \leq \widehat{\beta} \leq 3.63 \; 10^{-4}$,  depend on the biological species, 
  essentially on the total \emph{exonic} $GC$ content. 
 The four constants have to satisfy the normalization condition
\begin{equation}
\sum_{n} \, f(n) = 1
  \end{equation}
 The value of the constant $\widehat{\gamma} = 0.0164$ is approximately equal to $1/61$, i.e. the value of the codon usage probability in the
case of uniform and not biased codon distribution (not taking into account the 3 Stop codons), so really eq.(\ref{eq:bf}) depends on only two free parameters. 
Therefore the first two
terms in eq.(\ref{eq:bf}) can be viewed as the effect of some bias mechanism. 
We assume that this bias is only  the effect of the mutation  
and selection pressure, which we modelise by the effect of a suitable fitness and a mutation matrix
which depend on the change of the labels identifying the codons in the so called {\bf crystal basis} model  of the genetic code, see  (Frappat, Sciarrino and Sorba, 1998).  
The paper is organised in the following way. In Sec. 2 we briefly review  the mathematical
tools we use,  putting in an Appendix, to  make the article self-consistent, the basic definitions and properties. We identify a sequence of $N$ nucleotides or a $N$ spins chain  as a vector state of an irreducible representation (irrep.) of $U_{q \to 0}(sl(2))$.  Transitions between sequences are introduced
in terms of operators connecting vector states belonging or not belonging to the same irrep..  In Sec. 3 we build up a quantum spin model described by a hamiltonian whose diagonal
part, in the  basis vectors of the  irrep., represents the {\it fitness} and the off diagonal terms describes the
mutations. Let us point out that we do not aim to describe mutations in DNA as quantum effects. We use
the quantum mechanics  formalism only as a very useful language to introduce the mutations inducing
operators. The  model, which can appear unphysical if applied to a 
  quantum spin chain, should be considered, on the light of the previous 
  remarks on the application to the biological evolution,   as a guideline toward
   the search of solutions which can reproduce the observed 
  oligonucleotide distribution.  In some sense
we proceed in the backward direction with respect to the usual approach: we go from the quantum
 to the classical model. In Sec. 4, using the results of the previous section, we write classical kinetic equations for the probabilities and we solve it numerically, in the case of short oligonucleotide sequences. In Sec. 5 we discuss our results.  In Sec. 6 we extend the model to four letter alphabet, that is we  identify the nucleotides  with the fundamental 4-dim irreducible representation 
 of $U_{q \to 0}(sl(2) \oplus sl(2))$.  In Sec. 7 the four letters model described in Sec. 6 is applied
 and numerically solved for the codons.  
 The numerical solution of
     the model gives a stationary configuration for the distribution frequency
     which is indeed nicely fitted by the function $f(n)$. These solutions,
     but largely not their shape, depend on the numerical values of the 
     arbitrarily choosen parameters of the mutation matrix and of the fitness.
     However  a choice of the parameters in severe contradiction with the 
     reality, implying, e.g.,  a ratio of transversions over transitions 
     mutations very high or very low, seems to destroy the goodness of the fit.
 At the end a few conclusions and future  possible developments
 are presented. 

 \section{Mutations and Crystal basis}

An ordered sequence of N nucleotides, characterized only by the purine or 
pyrimidine character,  that is a string of $N$ binary labels or spins, can be represented as a vector state belonging to the 
N-fold tensor product of the fundamental irriducible representation 
(irrep.) (labeled by $ J =1/2$) of $\mathcal{U}_{q \to 0}(sl(2))$ 
 (Kashiwara, 1990), see 
Appendix A. This parametrization allows to represent, in a simple way,
the mutation of a  sequence as a transition between vectors, which
can be subjected to selection rules and  
whose strength  depends from the two concerned states. 
 
 \subsection{Labelling the state}
 
We identify a N-nucleotide sequence as a state 
\be
\mid \mathbf{J} \rangle = \mid J_3, J^N, \ldots, J^2 \rangle
\ee
where $\JN$ labels the irrep. which the state belongs to, $J_3$ is the value of the 3rd
diagonal generator of $\mathcal{U}_{q \to 0}(sl(2))$ ($2J_3 = n_R - n_Y$, $n_X$ being the
number of $X$ elements in the sequence) and $J^i$ ($2 \leq \; i  \; \leq N - 1$) are $ N - 2 $ labels
needed to remove the degeneracy of the irreps. in the $N$-fold tensor product in order to
completely identify the state. These further labels can be seen as the labels identifying
the irrep. which the state, corresponding to the sequence truncated to the $i$-th element,
belongs to.
We introduce a scalar product, such that
\be
\langle \mathbf{J}\mid \mathbf{K} \rangle =
\left\{
\begin{array}{rl}
1 & \mbox{if } J_3 = K_3 \mbox{ and } J^i = K^i \; \forall i \\
0 & \mbox{otherwise}
\end{array}
\right.
\ee
As an example, we can consider a trinucleotidic string
($N=3$) and label the eight different spin chains in the following way,
using the crystal basis representation $\mid {J_3},{J^N},\ldots,{J^2}\rangle$: 

\begin{eqnarray*}
\upa\doa\doa &=& \mid -\half,\half,0 \rangle  \;\;\;\;\;\;\;\; 
\upa\doa\upa = \mid \half,\half,0 \rangle \\
\doa\upa\doa &=& \mid -\half,\half,1 \rangle  \;\;\;\;\;\;\;\; 
\upa\upa\doa = \mid \half,\half,1 \rangle \\
\doa\doa\doa &=& \mid -\thalf,\thalf,1 \rangle \;\;\;\;\;\;\;\; 
\doa\doa\upa = \mid -\half,\thalf,1 \rangle \\
\doa\upa\upa &=& \mid \half,\thalf,1 \rangle \;\;\;\;\;\;\;\;\;\;\;
\upa\upa\upa = \mid \thalf,\thalf,1 \rangle.
\end{eqnarray*}

In our approach, sequences with the same number of 
spin up and down, placed in different sites, are described by different 
states. This has a phenomelogical support; indeed, in the case of RNA sequences,
the values of the free energy ($- \Delta G$),  on Kcal/mol at standard conditions, 
for four different sequences made of two CC  and two GG, i.e. two R 
and two Y, as reported in Table I
of (Xia et al., 1998):  CCGG (4.55), GGCC (5.37), CGCG (3.66), GCGC 
(4.61), are different. 
At this stage the crystal basis  provides, at least. an alternative way of labelling any 
finite spin sequence, mapping any sequence in a vector state of an 
irrep., but we know that in physics and mathematics the 
choice of appropriate variables is of primary importance to face a 
problem. Indeed we argue that these variables are suitable to 
partially describe non local events which affect the mutations.
We only consider  a single spin flip, which in most cases, but not 
always, is equivalent to a single nucleotide mutation. 
Flipping one spin can induce a transition to a state belonging or not 
belonging to the irrep of the original state. From the results of 
Appendix A we see that to identify a nucleotide sequence as a state of 
an irrep. requires to fix the number of RY contracted couples occurring 
in the considered sequence \footnote{For readers familiar with physics formalism, contraction should be 
 understood in the same sense of contraction of creation-annihilation 
 operators in the Wick expansion}. 
   Therefore, flipping a spin implies or the creation or the deletion of 
   a RY contracted couple, corresponding respectively to a variation 
   of -1 o +1 on the value  of the $J^N$   and, in case, of some others $J^i$
($2 \leq i \leq N-1$), or to leave unmodified the number of 
contracted couples (so that the variation of $J^{N}$ is $\Delta{J^{N}}=0$, but some other
$J^{i}$  may change).
In the following we classify the mutations of a single spin flip in
 N-nucleotide string, according to the induced variation in 
 the string labels $J_{3}, J^{N},\ldots,J^{2}$. We focus our attention on the spin flip
 of the $i$-th position,  but sometimes the transition will also effect other nucleotides.
We call \textit{left} (\textit{right}) \textit{side free}  the nucleotides
on the left (right) of $i$-th position and not contracted (in the sense expressed in Appendix A) with
another one on the same side. 
Let $R_l$ be the initial (before mutation) number of the \textit{left side free} purines and $Y_r$
the initial number of the \textit{right side free} pyrimidines.
We want to count the total number of contracted $RY$ couples (before and after mutation) in the
string, so we call $R_{in}$ ($R_{fi}$) the number, in the initial (final) state, of $R$ preceding
some $Y$, which is not on the same side, and not contracted with any $Y$ on their side. In the same way, with $Y_{in}$ ($Y_{fi}$)
we refer to the number of $Y$ following some $R$, which is not on the same side, and not contracted with any $R$ on their side.
If a $R \rightarrow Y$ mutation ($\Delta{J_3}=-1$) occurs in $i$-th position, then $R_{in}=R_{fi}+1$
and $Y_{in}=Y_{fi}-1$, where $R_{in}=R_{l}+1$ and $Y_{in}=Y_{r}$. We can distinguish different string
configurations around the $i$-th position, so that a single nucleotide mutation in $i$-th position
can correspond to different variations in the string labels. We have that
$\Delta{J^N} = |{R_{fi}}-{Y_{fi}}|-|{R_{in}}-{Y_{in}}|$

\begin{itemize}

\item If $\mathbf{R_l}=\mathbf{Y_r}$ then $R_{in}-1=Y_{in}$, so that $|R_{in}-Y_{in}|=1$; after mutation, $R_{fi}=Y_{fi}-1$, so that $|R_{fi}-Y_{fi}|=1$. Then the
variation of $J^{N}$ is $\Delta{J^{N}}=0$. We distinguish two subcases:

\begin{enumerate}

\item $R_{l}=Y_{r}\neq{0}$: $\jdue=0,\ldots,\jim=0,\ji=-1,\ldots,\jkm=-1,\jk=0,\ldots,\jn=0$
      ($2 \leq N-1; i+1 \leq k \leq N$);

\item $R_{l}=Y_{r}={0}$: $\ji=0 \; \forall{i}$.

\end{enumerate}

\item If $\mathbf{R_l}>\mathbf{Y_r}$, i.e. $R_{l}=Y_{r}+g$ ($g>0$), then $|R_{in}-Y_{in}|=g+1$ and
$J^{N}=\frac{1}{2}(g+1)$; after mutation, $|R_{fi}-Y_{fi}|=g-1$ and $J^{N}=\frac{1}{2}(g-1)$. Then
$\jn=-1$. We distinguish two subcases:

\begin{enumerate}

\item $Y_{r}=0$: $\jdue=0,\ldots,\jim=0,\ji=-1,\ldots,\jn=-1$ ($2 \leq i \leq N$);

\item $Y_{r}\neq{0}$: $\jdue=0,\ldots,\jim=0,\ji=-1,\ldots,\jn=-1$ ($3 \leq i \leq N-1$).

\end{enumerate}

\item If $\mathbf{R_l}<\mathbf{Y_r}$, i.e. $R_{l}=Y_{r}-g$ ($g>0$), then $J^{N}=\frac{1}{2}(g-1)$;
after mutation, $J^{N}=\frac{1}{2}(g+1)$, so that $\jn=1$. We distinguish two subcases:

\begin{enumerate}

\item $R_{l}=0$: $\jdue=0,\ldots,\Delta{J^{m-1}}=0,\Delta{J^{m}}=1,\ldots,\jn=1$ ($2 \leq m \leq N$,
$m\neq{i}$);

\item $R_{l}\neq0$:$\jdue=0,\ldots,\jim=0,\ji=-1,\ldots,\jkm=-1,\jk=0,\Delta{J_{k+1}}=1,\jn=1$
($2 \leq N-2; i+1 \leq k \leq N-1$).

\end{enumerate}

\end{itemize}

In the case of mutation $Y \rightarrow R$, for a fixed string configuration, the selection rules
are similar, changing ${\pm}1$ with ${\mp}1$.
Operators which lead to the above transitions can be built by $\jm,\ai,\aik$ and their adjoint
operators, defined in the following section.            

\subsection{Transition operators}

In this section we  write  the transition part of the hamiltonian, for the 
different possible initial configurations of the string.
  We  distinguish different string configurations around the $i$-th 
position, so that a single nucleotide mutation in $i$-th position can 
correspond to different variations in the string labels. 
The transitions inducing operators  are built by means of $\jm,\ai,\aik$ and 
their adjoint operators, as below defined.   

\begin{itemize}

\item If $\mathbf{R_l}=\mathbf{Y_r}. \;$ We distinguish two subcases:

\begin{enumerate}

\item $R_{l}=Y_{r} \neq {0}$
\be 
{H_1}=\s{i}{2}{N-1}\s{k}{i+1}{N} \, \alpha_{1}^{ik} \,( {\aik\jm + \jp\aikd})
\label{eq:1}
\ee
 \item $R_{l}=Y_{r}={0}$
 \be
{H_2}=  \alpha_{2} \, (\jm+\jp) \label{eq:2}
\ee
\end{enumerate}
\item If $\mathbf{R_l}>\mathbf{Y_r}. \;$
We distinguish two subcases:
\begin{enumerate}
\item $Y_{r}=0$ 
\be
{H_3}=\s{i}{2}{N} \, \alpha_{3}^{i} \, ({\ai\jm + \jp\aid}) \label{eq:3}
\ee
\item $Y_{r}\neq{0}$
\be
{H_4}=\s{i}{3}{N-1} \, \alpha_{4}^{i} \, ({\ai\jm + \jp\aid}) \label{eq:4}
\ee
\end{enumerate}
\item If $\mathbf{R_l}<\mathbf{Y_r}. \;$
We distinguish two subcases:
\begin{enumerate}
\item $R_{l}=0$
 \be
{H_5}=\s{m}{2}{N} \, \alpha_{5}^{m} \,({\jm A_{m}^{\dagger} + A_{m}\jp}) 
\label{eq:5}
\ee
\item $R_{l}\neq0$
\be
{H_6}=\s{i}{2}{N-2}\s{k}{i+1}{N-1} \, \alpha_{6}^{ik} \,
({\aik \jm A_{k+1}^{\dagger} + \aikd A_{k+1} \jp})  \label{eq:6}
\ee
\end{enumerate}
\end{itemize} 
 where  $\jp$ and $\jm$ are the \textit{step operators} defined by Kashiwara
(Kashiwara, 1990), acting on an
irreducible representation with highest weight $J^{N}$,  i.e. 
inducing 
the transitions $\Delta J^i=0, \; \forall i$  
\bea
A_{i,k} \mid \mathbf{J} \rangle &=& \mid J_3, J^N,.,
	J^k, J^{k-1}-1,., J^{i}-1, J^{i-1},., J^2 \rangle 
 \nonumber \\     
& & (2 \leq i \leq N-1 \;\;\;\; i+1 \leq k \leq N)  \label{eq:ik} 
\eea
\bea
A_i \mid \mathbf{J} \rangle &=& \mid J_3, J^{N}-1, \ldots,
    J^{i}-1, J^{i-1}, \ldots, J^2 \rangle
 \nonumber \\
& & (2 \leq i \leq N)
\eea
 \bea
B_m \mid \mathbf{J}\rangle &=& \mid J_3, J^{N}+1, \ldots,
    J^{m}+1, J^{m-1}, \ldots, J^2 \rangle
    \nonumber \\
    & & (2 \leq m \leq N)
 \eea
Therefore $\aikd$ is the operator which increase by 1 the value 
of $J^l$, for $k-1 \leq l \leq i$ and $B_m = A_m^{\dag}$.
Let us remark that in the above equations only the writing order of  
$A_{k+1}$ and $J_{\pm}$ has to be respected as
\be
[\, A_{k+1}, \; J_{\pm}] \neq 0
\ee
while ($ i < k < N$)
\be
[\, A_{i,k}, \; A_{k+1}] = [\, A_{i,k}, \; J_{\pm}] = 0
\ee
The following commutation relations can be useful for 
understanding the action of the transition hamiltonian as well as for further 
developments:
\be
[\, A_{i}, \; J_3] = [\, A_{i,k}, \; J_3] = 0 \;\;\;\;\; \forall i, k
\ee
\be
[\, A_{i} J_{-}, \; J_3] = A_{i} J_3 \;\;\;\;\;
[\, J_{+} A^{\dag}_{i}, \; J_3] = - J_{+} A^{\dag}_{i} 
\ee
\be
[\, A_{i,k} J_{-}, \; J_3] =  A_{i,k} J_{-} \;\;\;\;\;
[\,J_{+} A^{\dag}_{i,k}, \; J_3] = - J_{+}  A_{i,k} 
\ee
\be
[\, A_{i,k} J_{-} A^{\dag}_{k+1}, \; J_3] = A_{i,k} J_{-} A^{\dag}_{k+1}
\ee
\be
[\, A^{\dag}_{i,k}  A_{k+1} J_{+} , \; J_3] = - A^{\dag}_{i,k}  A_{k+1} J_{+}
\ee

A few words to comment on the above equations. Let us consider a mutation
$R \rightarrow Y$ which involve a transition
$\JN=-1$ (case $R_{l}>Y_{r}$);  the considered transition also entails
$\Delta{J_3}=-1$, so we
have to apply the operator $\jm$, as well as the operator $\ai$. Of course, first we have to lower by
1 the value of $J_3$, then to modify $J^N$, otherwise the initial state may 
eventually be annihilated, even if the
transition is allowed (in the case $J^{N}-1<J_3$).
Likewise, in corrispondence of a transition $Y \rightarrow R$ ($\Delta{J_3}=+1$), first the change
$J^{N} \rightarrow J^{N}+1$ has to be maked, then $J_3 \rightarrow J_{3}+1$.
   To write a self-adjoint operator, we have to add to the operator, which gives 
   rise to the
transition $Y \rightarrow R$,  the one which leads to $R \rightarrow Y$, leaving the rest of the
string unmodified, that is
\be
A_{i}J_- + {J_+}{A_{i}^{\dagger}}
\ee
This operator leads to the mutation $Y \rightarrow R$ or $R \rightarrow Y$ for a nucleotide in
$i$-th position, in a string with $R_{l}>Y_{r}$.
If the mutation $R \rightarrow Y$  rises the value of $J^{N}$, 
first $J^{N}$ has to be modified, then $J_3$;
with the aim to write a self adjoint operator, we  write
\be
  {J_-}{A_{m}^{\dagger}} + {A_m}J_+
\ee
 The above operator gives rise to mutations $R \rightarrow Y$ and
 $Y \rightarrow R$
for a nucleotide in $i$-th position, preceding the $m$-th one, in the case $R_{l}=0, Y_{r}\neq{0}$.  
Let us remark that eq.(\ref{eq:4}) is included in  
eq.(\ref{eq:3}), if the coupling constants $\alpha_{4}^{i}$ are assumed  equal
to $\alpha_{3}^{i}$;
 in  eq.(\ref{eq:6}), only  the writing order for  $A_{k+1}$ (and its adjoint)
and $J_{\pm}$ has to be respected.
 Let us also note that when $\Delta \JN = 0$ there is no need to order the operators.
 
\section{The quantum spin model}

Assuming now that the coupling constants  do not depend on $i,k,m$, we can write  the transition hamiltonian $H_I$ as 
 \be
{H_I}=\mu_{1}({H_3}+{H_5})+\mu_{2}{H_1}+\mu_{3}{H_2}+\mu_{4}{H_6}
\ee
The total hamiltonian of the model will be written as
\be
 H = H_{0} \, + \, H_I
\label{eq:H}
\ee
where $ H_{0}$ is the diagonal part in the choosen basis and,  
in the following, is assumed to be $H_{0} =  \mu_{0} \,  J_{3}$.
We let the fenomenology suggests us the scale of the values of the 
coupling constants of $H_I$. We want to write an interaction
term which makes the mutation in alternating purinic/pyrimidinic tracts less
likely than in
polypurinic or polypyrimidinic ones. We mean as a single nucleotide mutation in a polypurinic
(polypyrimidinic) tract, a mutation \emph{inside} a string with all nucleotides $R$ ($Y$), i.e. 
a highest (lowest) weight state. Such a transition corresponds to the 
selection rules
$\Delta J^N=-1$,$\Delta{J_3}=\pm1$, i.e. a transition generated by the action of $H_3$ and $H_5$.
In the interaction term $H_I$, we give them a coupling constant smaller than the 
others terms.    
We introduce, for  $ \; \Delta J_{3} = \pm 1 \;$, only four
different mutation parameters $\;\mu_{i}\;$ ($i = 1,2,3,4$),  with 
$\;\mu_{1}  < \mu_{k},\;\; k > 1$. 
\begin{enumerate}
\item $\mu_{1}$ for mutations which change the irrep., $ \; \Delta {J^N} = \pm 
1, \; $and include the spin flip inside an highest or lowest weight vector; 
\item $\mu_{2}$ for mutations which do not change the irrep., $ \; \Delta {J^N} = 0, 
 \; $ but modifies other values of $J^{k}$,  $\; \Delta J^{k} = \pm 1$;
\item $\mu_{3}$ for mutations which do not change the irrep., $ \; \Delta J^{N} = 0$,
      neither the other values of $J^{k}$, $\; \Delta J^{k} = 0$, ($2 \leq k \leq N-1$);  
\item $\mu_{4}$ for mutations which change the irrep., $ \; \Delta {J^N} = \pm 
1, \; $ but only in a string with $0 \neq R_{l} < Y_{r}$.
\end{enumerate}
 We do not introduce another parameter, for mutations
generated by $H_4$, i.e. $i$-th nucleotide mutation in a string with
$R_{l} > Y_{r} \neq 0$,  not to distinguish, in a polypurinic string, between 
a mutation in $2$-th position and another one inside the string.
Let us emphasize once more that the proposed model takes into account, 
at least partially, the effect on the transition in the $i$-th site of 
the distribution of all the spins.
Presently we consider  only  the  part of the interaction hamiltonian $H_I$, which generates transitions corresponding to one spin flip write, but, in analogous way,  we could write more complicated 
transition operators.
 Let us illustrate, in a simple example, the difference between this 
scheme and the standard one, based on the hypotesis of transition
probability between chains, only depending by their Hamming distance. Let us 
consider the following string $RRRRR$. By a single flip spin  the string 
goes in one of the following configurations:
\be
 1) \; YRRRR \;\;\; 2) \; RYRRR \;\;\; 3) \; RRYRR  \;\;\;
 4) \; RRRYR \;\;\; 5) \; RRRRY \label{eq:states}
\ee
In the models based on the Hamming distance, all the transitions are 
equally probable, as the final strings are all at the same distance from 
the original one. In the present scheme the 1-rst transition is ruled by the 
value 
of $\; \mu_{3}$, the transition 2-5 are ruled by $\; \mu_{1} \;$.

Let us stress that our scheme is not equivalent to an Ising model 
 with   the transition strength depending on the position. To 
illustrate the difference with  a few examples let us consider the 
transitions $\; RYYRR \to RYYYR\;$, $\;RYYRY \to RYYYY\;$, both
with a flip in the fourth position, the first one ruled by
$\mu_{2}$, the second one by $\mu_{1}$. Mutations in different points
can be ruled by the same coupling constant:  $\; RYYRR \to RYYYR\; $,
$\; RYRRY \to RYYRY\; $ 
with a flip, respectively, in the 
4th and 3rd position ruled by $\; \mu_{2}$. 
As already said, the main motivation for introducing this quantum model is that it provides
 the formal and conceptual language to  write  the transitions, ensuring in the same time, due to the unitary character of the evolution operator,  the conservation of the probability.  We shall briefly describe in Sec. 6  the outcome of this model, see (Minichini and Sciarrino 2004a) for more details,  which has only been reported to make, hopefully, more clear the structure of the classical model of the next section.  Let us point out that  there are very strong drawback in trying to further pursue the study of the quantum model, for example
   superposed states, that is linear combinations of sequences, do exist in  such models, while
 only the different sequences have a biophysical interpretation.
 
\section{The classical model}

In the previous section we have introduced mutation inducing operators based on the change of the global labels $J^{i}$. Using these results as a guide we write a kinetic equations systems in which
the non vanishing mutation matrix entries depend on the labels of the connected sequences.
We are interested in finding the stationary or equilibrium configuration of the $2^{N}$
different possible 
 sequence. Writing $p_{\mathbf{J}}(t)$ the probability distribution at 
 time $t$ of the sequence identified by the vector $\mid \mathbf{J} 
 \rangle$, a decoupled version of selection mutation equation, (see
(Hofbauer and Sigmund, 1988)  for an exhaustive review), for a haploid organism, can be written as
 \be
 \frac{d}{dt} \, p_{\mathbf{J}}(t) =  p_{\mathbf{J}}(t)\left(R_{\mathbf{J}}-
\sum_{\mathbf{K}}\;R_{\mathbf{K}}\; p_{\mathbf{K}}(t)\right)+\sum_{\mathbf{K}}\;M_{\mathbf{J,K}}\; p_{\mathbf{K}}(t)
 \label{eq:ME}
\ee 
where $R_{\mathbf{K}}$ is the Malthusian fitness of the sequence corresponding to
the vector $\mid \mathbf{K} \rangle$ and $M_{\mathbf{J,K}}$ are the entries of a
mutation matrix $M$ which satisfies 
\be
{\Large M}_{\mathbf{J},\mathbf{J}} = - \, \sum_{\mathbf{K} \neq 
\mathbf{J}} \; {\Large M}_{\mathbf{J},\mathbf{K}}
\label{eq:norm}
\ee
The equation (\ref{eq:ME}) is reduced to
\be
\frac{d}{dt} \, x_{\mathbf{J}}(t) = \sum_{\mathbf{K}}\;\left(H+M\right)_{\mathbf{J,K}}
\;x_{\mathbf{K}}(t)
\label{eq:Hamilt}
\ee
where
\be
x_{\mathbf{J}}(t) =  p_{\mathbf{J}}(t)\exp\left(\sum_{\mathbf{K}}\;R_{\mathbf{K}}\;
\int_{0}^{t}\; p_{\mathbf{K}}(\tau)\,d\tau \right)
\label{eq:transf}
\ee
and $H$ is a diagonal matrix, with fitness as entries ($R_{\mathbf{K}}  = H_{{\mathbf{K}},{\mathbf{K}}}$).
In our model the mutation matrix is written as the sum of the 
 partial mutation matrices  ${M_i}$ which are obtained by  the
 interaction hamiltonians  ${H_i}$ replacing the adjoint operators by the transposed
 (denoted by an upper labe $^T$).
 Assuming now that the coupling constants do not depend on $i,k,m$, 
we can write the mutation matrix $M$ as  (Minichini and Sciarrino, 2004b)
 \be
{M}=\mu_{1}({M_3}+{M_5})+\mu_{2}{M_1}+\mu_{3}{M_2}+\mu_{4}{M_6} + M_{D}
\ee
where $M_{D}$ is the diagonal part of the mutation matrix defined by 
eq.(\ref{eq:norm}).
The hierarchy of the values of the coupling constants is fixed as in the previous
section.

 \section{Results}
 
\begin{figure}[tbh]
\begin{center}
\framebox{\epsfxsize=0.4\textwidth
\epsffile{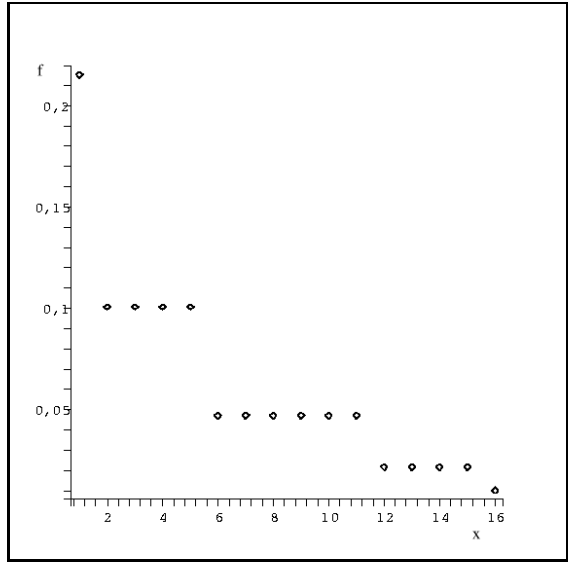}}
\end{center}
\caption{Rank ordered distribution of equilibrium population (N=4) obtained for a Hamming 
transition matrix, with 
$\alpha=0.60$.}
\label{stepH16}
 \end{figure}

\begin{figure}[tbh]
\begin{center}
\framebox{\epsfxsize=0.4\textwidth
\epsffile{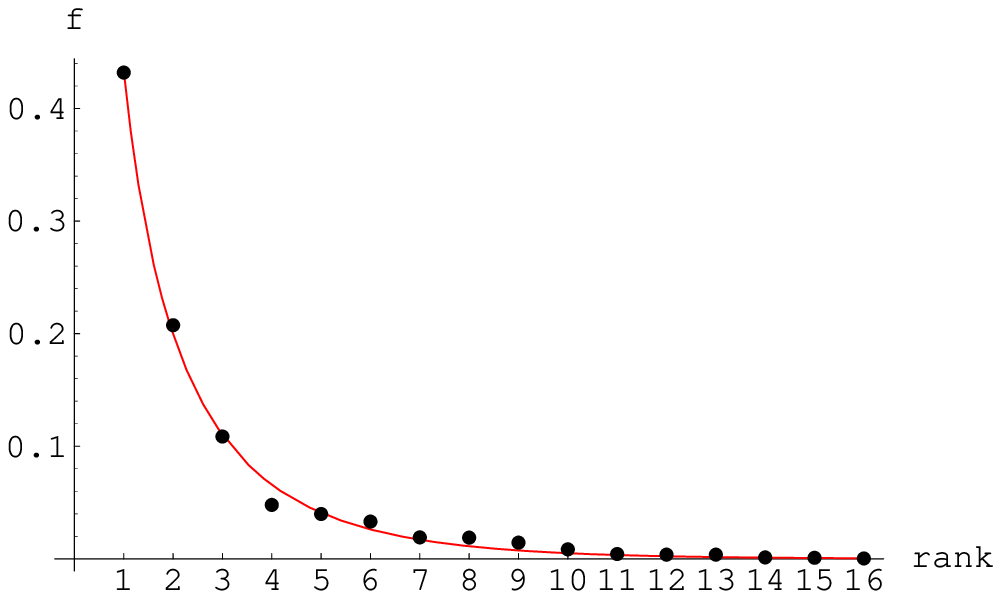}}
\end{center}
\caption{Rank ordered distribution of equilibrium population (N=4) for a 
transition matrix $M$ with 
$\epsilon=0.25,\;\gamma=\delta=\eta=0.50$.
The distribution was fitted by a Yule function (continuous line) 
$f=a{R^k}{b^R}$ ($R$ is the
rank). The parameters were estimated as $a=0.37,\,b=1.02,\,k=-1.28$.}
\label{YuleH16}
 \end{figure}

\begin{figure}[tbh]
\begin{center}
\framebox{\epsfxsize=0.4\textwidth
\epsffile{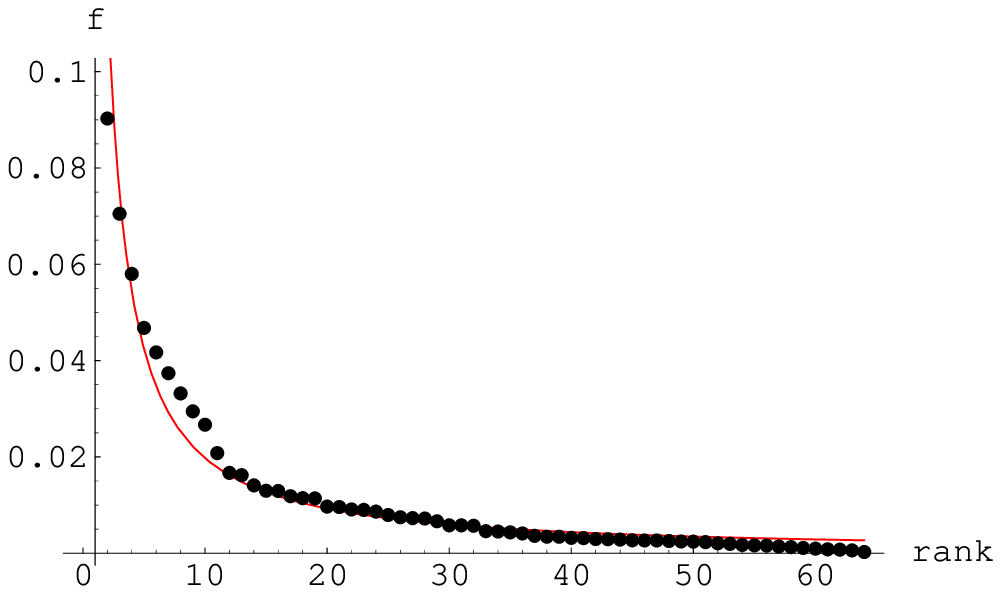}}
\end{center}
\caption{Rank ordered distribution of equilibrium population (N=6) for a transition matrix 
$M$, with
$\epsilon=0.25,\;\gamma=\delta=\eta=0.50$. The distribution was fitted 
by a Yule
function (continuous line) $f=a{R^k}{b^R}$ ($R$ is the rank).  The 
parameters were estimated as $a=0.26,\,b=1.00\,k=-1.11$.} 
\label{N=6}
 \end{figure}

 The  evolution equation of the model for the probabilities will be written in 
terms of the matrix  $\bar{H}= H + M + \lambda\mathbf{1}$,
where the fitness can be $H=J_3$ (purely additive fitness)
 and $\lambda$ is choosen in such a way to guarantee $\bar{H}$ is
positive. Being $H + M$ irreducible, the composition of equilibrium population is
given by
\be
p_{\mathbf{J}} = 
\frac{\tilde{x}_{\mathbf{J}}}{\sum_{\mathbf{K}}\;\tilde{x}_{\mathbf{K}}}
\ee
where $\tilde{x}_{\mathbf{J}}$ is the Perron-Frobenius eigenvector, see, e.g.  (Encyclopedic Dictionary of Mathematics, 1960) of
$\bar{H}$. 
In  (Minichini and Sciarrino, 2004b) the
 numerical solutions of the model have been reported, with a suitable choice of the value of 
the parameters, for N = 3,4,6.  %
Before discussing these results, we point out
 explicitly the main features of our model. 
 $M$
 describes an interaction on the $i$-th spin neither depending on the
 position nor on the nature of the closest neighbours, but which  
takes into account, at least partially, the effects, on the transition 
in the $i$-th site, of the 
distribution of all the spins, that is non local effects. 
 Indeed it
 depends on the ``ordered" spin orientation surplus on the left and on
 the right of the $i$-th position. Should it not depend on the order, 
it
 may be considered as a mean-field like effect. Moreover $\Delta J_3 =
 \pm 1$ transitions   are allowed, which, e.g. for N = 4, can be considered or as the
 flip of a spin combined with an exchange of the two, oppositely
 oriented, previous or following spins or as the collective flip of
 particular three spin systems, containing a two spin system with
opposite spin orientations (see example below).
Biologically, the transition depends in some way on the  "ordered"
 purine surplus on the left and on the right of the mutant position.
 Let us briefly comment on the physical-biological meaning of the
 ``ordered" spin sequence. Our aim is to study finite oligonucleotide 
 sequence in which a beginning and an end are defined. This implies we 
 can neither make a thermodynamic limit on $N$ nor define periodic 
 conditions on the spin chain. So we have to take into account the 
 ``edge" or ``boundary" conditions on the finite sequence. An analogous 
 problem appears in determing thermodynamic properties of short 
 oligomers and, in this framework, in (Goldstein and Benight, 1992) the 
 concept of fictitious nucleotide pairs E and E' has been introduced, 
 in order to mimick the edge effects. The ordered couple of  RY takes 
 into account in some way the different interactions of R and Y with 
 the edges.
  For example, the transition matrix, on the above basis (for N = 3) 
is the following
one, up to a multiplicative dimensional factor $\mu_{0}$
\begin{equation}
\label{matrixModel}
M = 
\left(
\begin{array}{cccccccc}
x & \delta & 0 & \gamma & \epsilon & 0 & \epsilon & 0 \\
\delta & x & 0 & 0 & 0 & \epsilon & 0 & \epsilon \\
0 & 0 & x & \delta & \epsilon & 0 & \epsilon & 0 \\
\gamma & 0 & \delta & x & 0 & \epsilon & 0 & \epsilon \\
\epsilon & 0 & \epsilon & 0 & x & \delta & 0 & 0 \\
0 & \epsilon & 0 & \epsilon & \delta & x & \delta & 0 \\
\epsilon & 0 & \epsilon & 0 & 0 & \delta & x & \delta  \\
0 & \epsilon & 0 & \epsilon & 0 & 0 & \delta & x
\end{array}
\right)  
\end{equation}
where the diagonal entries $x$, not explicitly written, are given by eq.(\ref{eq:norm}).
Note that the above matrix depends only on three coupling 
constants
due to the very short length of the chain. For $N \ge 4$ the 4th 
coupling constant
(denoted in the following by $\eta$) will appear.
Let us emphasize that the mutation matrix $M$ (\ref{matrixModel}) does not 
only
connect states at unitary Hamming distance. As an example, we write 
explicitly the
transitions from $\mid \half,\half,0\rangle$ ($\upa\doa\upa$) 
and from
$\mid -\half,\half,0\rangle$ ($\upa\doa\doa$)
\begin{eqnarray*}
\upa\doa\upa \longrightarrow
\left\{
\begin{array}{c}
\upa\upa\upa \\
\doa\doa\upa \\
\upa\doa\doa
\end{array}
\right.& \qquad \upa\doa\doa \longrightarrow
\left\{
\begin{array}{c}
\doa\upa\upa \\
\doa\doa\doa \\
\upa\upa\doa \\
\upa\doa\upa
\end{array}
\right.
\end{eqnarray*}
The first transition of the second example can be regarded as a spin-flip of  the three spins.
 Let we explicitly write, for $N = 3$, the 
 mutation matrix, which  allows transitions only between chains at
Hamming distance equal to one, with coupling constant  $\alpha$. 
\begin{equation}
\label{matrixHamm}
H_\mathrm{Hamm} =
\left(
\begin{array}{cccccccc}
y & \alpha & 0 & \alpha & \alpha & 0 & 0 & 0 \\
\alpha & y & 0 & 0 & 0 & \alpha & 0 & \alpha \\
0 & 0 & y & \alpha & \alpha & 0 & \alpha & 0 \\
\alpha & 0 & \alpha & y & 0 & 0 & 0 & \alpha \\
\alpha & 0 & \alpha & 0 & y & \alpha & 0 & 0 \\
0 & \alpha & 0 & 0 & \alpha & y & \alpha & 0 \\
0 & 0 & \alpha & 0 & 0 & \alpha & y & \alpha  \\
0 & \alpha & 0 & \alpha & 0 & 0 & \alpha & y
\end{array}
\right)
\end{equation}
where the diagonal entries, not explicitly written, are given by eq.(\ref{eq:norm}).
 Note that, even if we put in eq.(\ref{matrixModel}) all the constant equal to $\alpha$
 ($\delta = \gamma = \varepsilon = \alpha$), we do not get the Hamming hamiltonian
 (\ref{matrixHamm}).
 If we order (in a decreasing way) the equilibrium probabilities, we obtain,
using the mutation matrix with Hamming distance, 
a rank
ordered distribution of transition probability like that in 
fig.\ref{stepH16} for $N=4$. Its 
shape does not depend on the value of $ \alpha$.
  The rank-ordered distribution 
of the probabilities shows a plateaux structure: 
every
plateaux contains spin sequences at the same Hamming distance from 
the  sequence with the highest value of the fitness.
Using the mutation matrix  (\ref{matrixModel}), the 
rank ordered probabilities distribution does not show a 
plateaux structure,
but its shape is well fitted by a Yule distribution 
(fig.\ref{YuleH16}), like the observed frequency  
distribution of oligonucleotidic in the strings of nucleic 
acids  (Martindale and Konopka, 1996). 
Let us observe that we obtain a Yule distribution (and not a plateaux 
structure)
even if all parameters in (\ref{matrixModel}) are tuned at the same 
value, which means that the distribution is the outcome of the model 
and not of the choice of the values of the coupling constants.
Analogous resultes are obtained for $N=6$ 
(fig.\ref{N=6}).
 Let us point out that:

i) our model is not equivalent to a model where
  the intensity  depends on the site  undergoing the 
  transition, or from the nature of the closest neighbours or the 
  number of the $R$ and $Y$ labels of the sequence; indeed 
essentially 
  the intensity depends on distribution in the sequence of 
  the $R$ and $Y$;
  
ii) the ranked distribution of the 
probabilities  follows  a Yule distribution law, but as the value of 
the parameter b is close to the unity, 
  the distribution is equally well 
fitted by a Zipf law (Zipf, 1949) ($f = a 
\, {n^k}$), in 
agreement with the remark of (Martindale and Konopka, 1996).

Let us  also briefly recall the outcomes of the  \textit{genetically inspired} quantum spin model presented in Sec. 3.
We can study the time evolution of an initial state, representing a given
spin chain, and evaluate the probability of transition in another one, if
$H$ is the hamiltonian which generates the dynamics of the system.
The matrix form of $H$, on the above basis  for a fixed initial state,
is obtained (for $N = 3$),  by replacing in eq.(\ref{matrixModel}) the diagonal terms  by  the eigenvalues of $J_{3}$, i.e. by, respectively, (-1,1,-1,1,-3,-1,1,3) (up a multiplicative factore $1/2$).
Analogously we can study the dynamics of an ordered quantum spin
chain, with  an interaction Hamiltonian, leading to transitions with the same probability
 between nucleotide strings at unit Hamming distance,  whose matrix, for $N = 3$, is obtained by eq.(\ref{matrixHamm}) by replacing the diagonal terms with the eigenvalues of $J_{3}$.
 In order to evaluate the probabilities of transition, we cannot analytically study
 the time evolution of an initial state, representing a fixed spin sequence, as ruled by
  eq.(\ref{matrixModel}) with the change of the diagonal terms, but we can find a numerical solution.   %
The transition probability between two states, belonging to the crystal basis,
 exhibits the quantum mechanically typical oscillating behaviour as a function of the time.
 We define a time-averaged transition probability (initial state (i) $ \lra $ final state (f))
 \be
 <p_{if}> \, = \, \frac{1}{T} \; \int_0^T \; p_{if}(t) \, dt         \label{eq:tav}
 \ee
 where the value of $T$   will be numerically fixed to  a value, such that the
 r.h.s.  of  eq.(\ref{eq:tav}) becomes stable.
If we order (in a decreasing way) the average transition probability from an 
initial state to every other chain, if (\ref{matrixHamm}) is the hamiltonian, we obtain a rank
ordered distribution of transition probability like that in fig.\ref{stepH16}. Its 
shape does not depend by the choice of initial state or by the coupling
constant $\alpha$ value.
We always get the same structure, for models with transition probability
only depending on Hamming distances. So the rank-ordered distribution
of the average transition probability shows a plateaux structure: every
step contains spin chains at the same Hamming distance from the initial one.
In the case of the model which we propose here, i.e. the hamiltonian in
(\ref{matrixModel}),  which we call crystal basis model, the distribution
of rank ordered average transition probability does not show a plateaux structure,
but its shape is well fitted by a Yule distribution like that in fig.\ref{YuleH16}.  
Also in the quantum model, we obtain a Yule distribution (and not a plateaux structure)
even if all parameters in (\ref{matrixModel}) are tuned at the same value.
 In this case, the state, labelled by 1 in the plots, is the initial one.  The ranked distribution of the 
probabilities, not averaged in time, computed for several values of 
the time, also follows generally a Yule distribution law. Moreover we still remark that,
for the highest value of $N$, the distribution is equally well 
fitted by a Zipf law, i.e. $ b = 1$ in eq.(\ref{eq:Yule}), but not for the lowest values of $N$, in 
agreement with the remark of  (Martindale and Konopka, 1996.

\section{The four letter model}

In order to label  a sequence of N nucleotides, taking into account that they belong to the four letter set \{C,T/U,G,A\}, we assign  the 4 nucleotides  to the 4-dim irreducible fundamental 
 representation (irreps.) $(1/2, 1/2)$ of $\mathcal{U}_{q \to 0}(sl(2) \oplus sl(2))$  (Frappat, Sciarrino and Sorba, 1998)  with the following assignment for the values of  the third component of
  $\vec{J}$ for the two $sl(2)$ which in the following will be denoted
   as  $sl_{H}(2) $ and $sl_{V}(2) $ :
\be 
	\mbox{C} \equiv (+\half,+\half) \qquad \mbox{T/U} \equiv (-\half,+\half) 
	\qquad \mbox{G} \equiv (+\half,-\half) \qquad \mbox{A} \equiv 
	(-\half,-\half)
	\label{eq:gc1}
\ee 
  It follows that an ordered sequence of N nucleotides can be represented as a vector belonging to the 
N-fold tensor product of the fundamental irriducible representation 
  of $\mathcal{U}_{q \to 0}(sl(2) \oplus sl(2))$,  in a straightforward generalization of the approach followe in Sec.2 for   $\mathcal{U}_{q \to 0}(sl(2))$.  In the following we use the symbols $X$ for C,G and $Z$ for U,A.  In the formalism of $\mathcal{U}_{q \to 0}(sl(2) \oplus sl(2))$     all the previous results have to be understood to refer to $sl_{V}(2)$.
 Now we identify a N-nucleotide sequence as a state 
\be
\mid \mathbf{J}_{H} \mathbf{J}_{V} \rangle = \mid J_{3,H}, J_{3,V}; J_{H} ^N, J_{V} ^N; \ldots ;J_{H} ^2, J_{V} ^2 \rangle
\ee
where $\JN_{m}$ ($m = H, V$) labels the irrep. which the state belongs to, $J_{3,m}$ is the value of the 3rd diagonal generator of $\mathcal{U}_{q \to 0}(sl_{m}(2))$ ($2J_{3,H} = n_X - n_Z$,  $2J_{3,V} = n_R - n_Y$) and $J_{m}^i$ ($2 \leq \; i  \; \leq N - 1$) are $ 2(N - 2) $ labels
needed to completely identify the state.   As an example,  the trinucleotidic string CGA is labeled by
\be
\mid  CGA \rangle  =  \mid \left( \half \right)_{H}, - \left(\half \right)_{V}; \left(\half \right)_{H}, \left(\half \right)_{V};   \left(1\right)_{H} \left(1 \right)_{V} \rangle 
\ee
The  previously introduced scalar product  is straightforwardly  
generalized. 
In the present paper,  we only consider  a single spin flip in H or V spin or in both H and V, which in most cases, but not always, is equivalent to a single nucleotide mutation.  Obviously a H spin flip (V and H,V  flip) corresponds, respectively,  to a biological \textit{transition}  (\textit{transversion}).
Flipping one spin can induce a transition to a state belonging or not 
belonging to the irrep. of the original state.   
  From an immediate generalisation of the results of 
Appendix A, we need, to identify a nucleotide sequence as a state of 
an irrep.,    to fix the number of RY and XZ contracted couples occurring 
in the considered sequence. 
   Therefore flipping a spin implies or the creation or the deletion of 
   a RY or XZ or both contracted couple, corresponding, respectively, to a variation 
   of -1 o +1 on the value  of the $J_{V} ^N$, $J_{H} ^N$  or both  and,  possibly, of some others $J_{m}^i$
($2 \leq i \leq N-1$), or to leave unmodified the number of 
contracted couples (so that   $\Delta J_{m} ^N = 0$, but some other
$J_{m} ^{i}$  are modified).
 We focus our attention on the spins flip
 of the $i$-th position and we go on in a completely analogous way as in Sec. 2,  but  taking into account  the two couples RY and XZ.  
Assuming, as previously,  that the coupling constants do not depend on $i,k,m$, 
we write the mutation matrix $M$ as 
 \bea
{M } & = & {M}_{H} + {M}_{V}  \nonumber  \\
& = & \mu_{1}({M_{3,H}} + {M_{5,H}}) + \mu_{2}{M_{1,H}} + \mu_{3}{M_{2,H}} + \mu_{4}{M_{6,H}} 
\nonumber \\
& + &  
\lambda_{1}({M_{3,V}} + {M_{5,V}}) + \lambda_{2}{M_{1,V}} + \lambda_{3}{M_{2,V}} + \lambda_{4}{M_{6,V}} + M_{D}
\label{eq:mhv}
\eea
where $M_{D}$ is the diagonal part of the mutation matrix defined by eq.(\ref{eq:norm}), and
${M_{k,m}}$ ($ k = 1,2,3,5,6; m = H,V$) are the off-diagonal mutation matrices defined by the
following operators, where we have omitted to explicitly write the coupling constants
\be
{H_{1,m}} =  {\aikm\jmm + \jpm\aikmd}
 \ee
 \be
{H_{2,m}} =   \jmm + \jpm 
\ee
 \be
{H_{3,m}} = {\aim\jmm + \jpm\aimd}  
\ee
 \be
{H_{4,m}} = {\aim\jmm + \jpm\aimd} 
\ee
 \be
{H_{5,m}} = {\jmm A_{m;m}^{\dagger} + A_{m;m}\jpm}
 \ee
 \be
{H_{6,m}} =  {\aikm \jmm A_{k+1;m}^{\dagger} + \aikmd A_{k+1;m} \jpm})  \label{eq:6b}
\ee
Note that in eq.(\ref{eq:mhv}) we have not introduced a coupling term between the two $sl(2)$, i.e. a mutation matrix of the type ${M}_{H,V} \propto \jph\jpv$  or ${M}_{H,V} \propto \jmh\jmv$.
In order to fit the phenomenological observation that the transitions occur more frequently than the transversions, we have to fix the coupling constants  $\lambda$ of the order of  $1/2 - 1/3$ of the coupling constants $\mu$. Let us remark that, with the chosen mutation matrix eq.(\ref{eq:mhv}), a single spin mutation does not correspond necessarily to a H-spin or V-spin flip. Indeed the mutations $ C \llra A$ amd $T \llra G$ imply a flip of both the H and V spins, therefore these mutations should be depressed.

\section{The rank ordered distribution of codons}

\begin{figure}[tbh]
\begin{center}
\framebox{\epsfxsize=0.4\textwidth
\epsffile{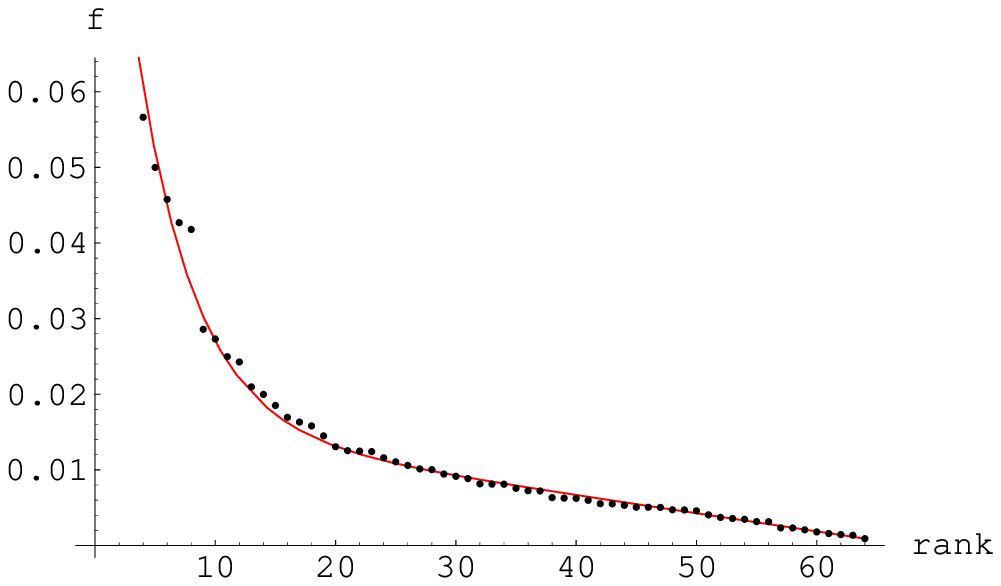}}
\end{center}
\caption{Rank ordered distribution of equilibrium population of codons, for the set of parameters ${\mu_{1,H}}=1.2,\,{\mu_{2,H}}={\mu_{3,H}}=3.6,\,{\mu_{1,V}}=0.4,\,{\mu_{2,V}}={\mu_{3,V}}=1.2$. 
The distribution was fitted by the function $f(n)=\alpha \exp(-\eta n) - \beta n +\gamma$ (continuous line), where $n$ is the rank of the codons. The parameters were estimated as $\alpha=0.084,\,\beta=2.77\cdot10^{-4},\,\gamma=0.018,\,\eta=0.219.$ }
\label{Par96}
 \end{figure}

\begin{figure}[tbh]
\begin{center}
\framebox{\epsfxsize=0.4\textwidth
\epsffile{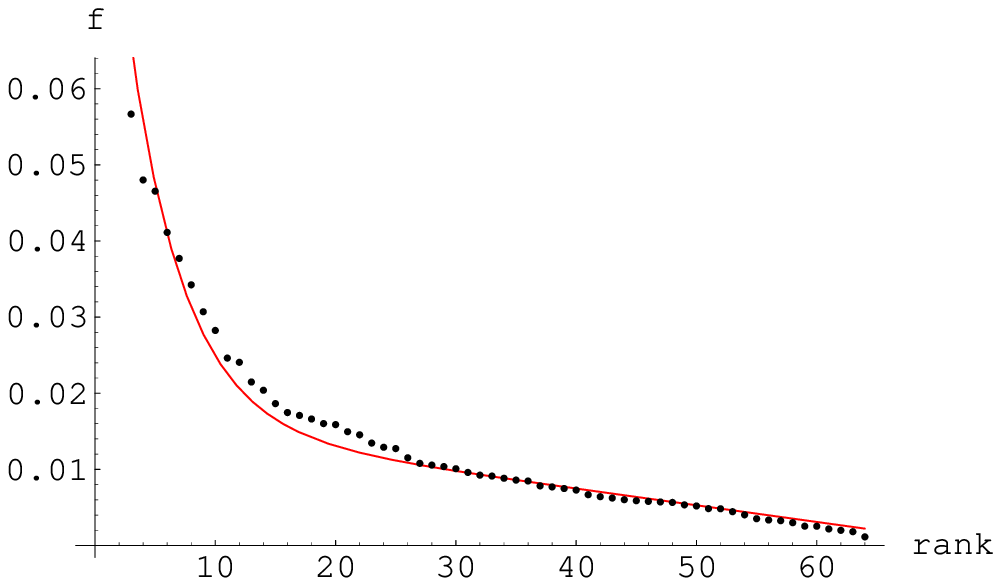}}
\end{center}
\caption{Rank ordered distribution of equilibrium population of codons, for the set of parameters
${\mu_{1,H}}=1,\,{\mu_{2,H}}={\mu_{3,H}}=3,\,{\mu_{1,V}}=0.5,\,{\mu_{2,V}}={\mu_{3,V}}=1.5$. 
The distribution was fitted by the function $f(n)=\alpha \exp(-\eta n) - \beta n +\gamma$ (continuous line), where $n$ is the rank of the codons.The parameters were estimated as $\alpha=0.076,\,\beta=3.46\cdot10^{-4},\,\gamma=0.022,\,\eta=0.285.$}
\label{Par99}
 \end{figure}

\begin{figure}[tbh]
\begin{center}
\framebox{\epsfxsize=0.4\textwidth
\epsffile{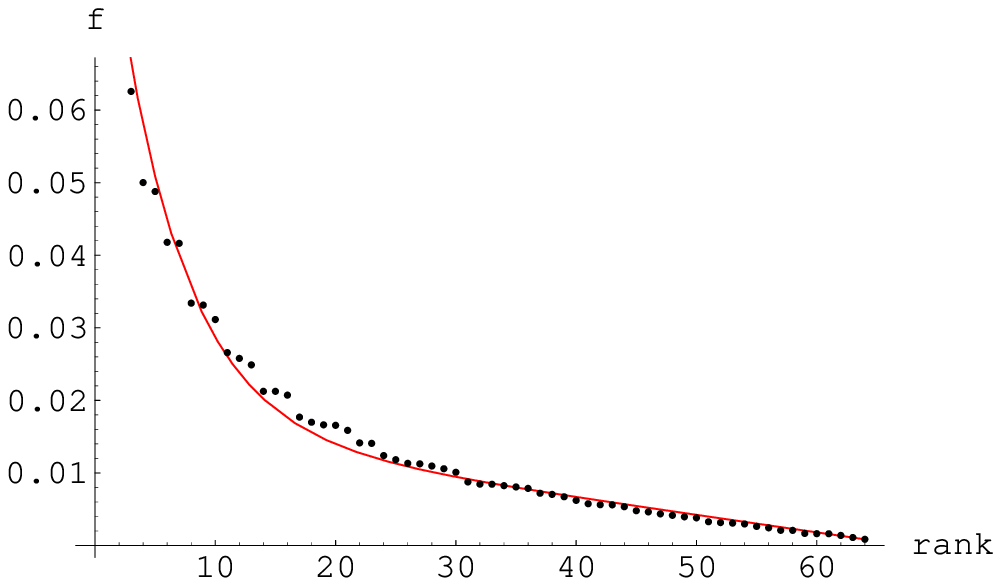}}
\end{center}
\caption{Rank ordered distribution of equilibrium population of codons, for the set of parameters 
${\mu_{1,H}}={\mu_{2,H}}=3,\,{\mu_{3,H}}=1,\,{\mu_{1,V}}={\mu_{2,V}}=1,\,{\mu_{3,V}}=0.3$. 
The distribution was fitted by the function $f(n)=\alpha \exp(-\eta n) - \beta n +\gamma$ (continuous line), where $n$ is the rank of the codons. The parameters were estimated as $\alpha=0.071,\,\beta=3.36\cdot10^{-4},\,\gamma=0.020,\,\eta=0.211.$}
\label{Par94}
 \end{figure}

\begin{figure}[tbh]
\begin{center}
\framebox{\epsfxsize=0.4\textwidth
\epsffile{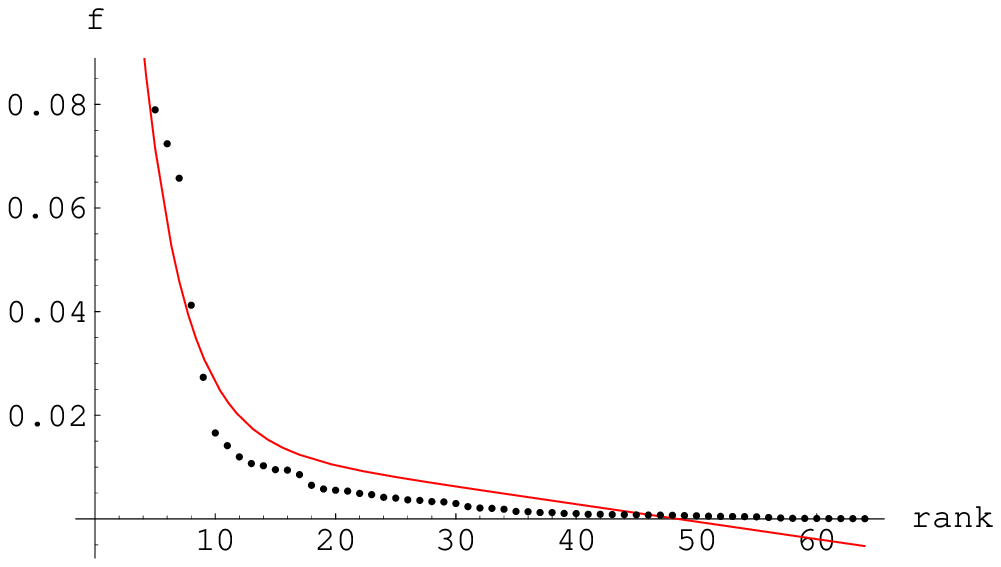}}
\end{center}
\caption{Rank ordered distribution of equilibrium population of codons, for the set of parameters 
${\mu_{1,H}}=1,\,{\mu_{2,H}}={\mu_{3,H}}=3,\,{\mu_{1,V}}=0.1,\,{\mu_{2,V}}={\mu_{3,V}}=0.3$. 
The distribution was fitted by the function $f(n)=\alpha \exp^{-\eta n} - \beta n +\gamma$ (continuous line), where $n$ is the rank of the codons.The parameters were estimated as $\alpha=0.187,\,\beta=8.75\cdot10^{-5},\,\gamma=0.005,\,\eta=0.246.$}  
\label{Par98}
 \end{figure}

\begin{figure}[tbh]
\begin{center}
\framebox{\epsfxsize=0.4\textwidth
\epsffile{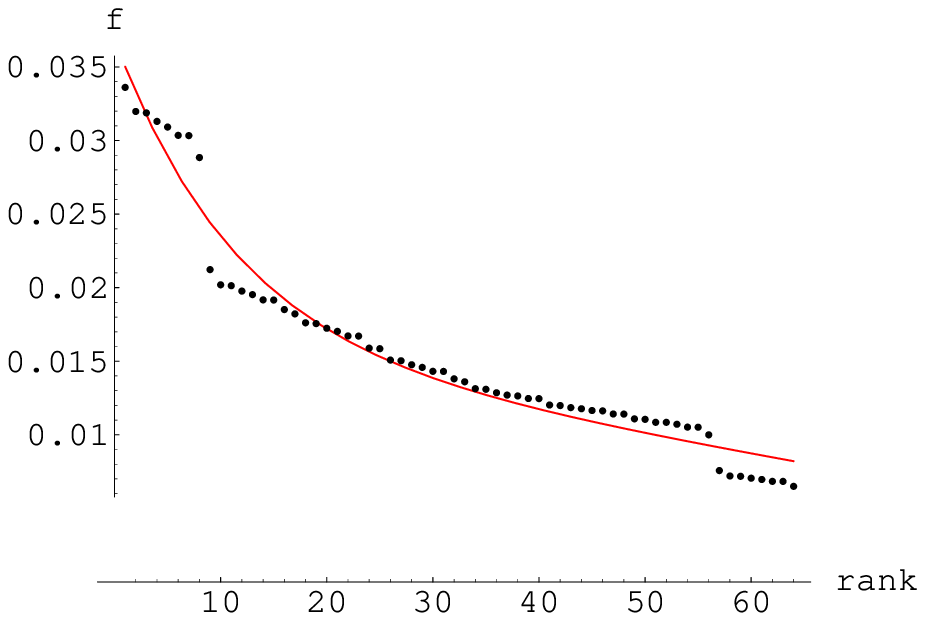}}
\end{center}
\caption{Rank ordered distribution of equilibrium population of codons, for the set of parameters 
${\mu_{1,H}}=1.1,\,{\mu_{2,H}}={\mu_{3,H}}=3.3,\,{\mu_{1,V}}=11,\,{\mu_{2,V}}={\mu_{3,V}}=33$. 
The distribution was fitted by the function $f(n)=\alpha \exp(-\eta n) - \beta n +\gamma$ (continuous line), where $n$ is the rank of the codons. The parameters were estimated as $\alpha=0.016,\,\beta=1.87\cdot10^{-4},\,\gamma=0.019,\,\eta=0.122.$ } 
\label{Par87}
 \end{figure}

In (Frappat, Sciarrino and Sorba, 1998) a mathematical  model, called crystal basis model, for the genetic code has been proposed where from the assignment eq.(\ref{eq:gc1}) of
  the four nucleotides    to the 4-dim fundamental 
$(\half,\half)$  irreducible representation of
the quantum group ${\cal U}_{q \to 0}(sl(2) \oplus sl(2))$,  the codons ($3$-nucleotide 
sequence) appear as composite state in the$3$-fold tensor product  of $(\half,\half)$.
 From the general formalism of the previous section, a codon is identified as a state
\be
\mid \mathbf{J}_{H}\rangle \, \otimes \mid \mathbf{J}_{V} \rangle\equiv 
  \mid \mathbf{J}_{H} \mathbf{J}_{V} \rangle = \mid J_{3,H}, J_{3,V}; 
J_{H} ^3 J_{V} ^3; J_{H} ^2, J_{V} ^2 \rangle
 \ee
For example we have, see (Frappat, Sciarrino and Sorba, 2001) for a list of all the states:
$$
\mid  CGA \rangle  =  \mid \left( \half \right)_{H}, - \left(\half 
\right)_{V}; \left(\half \right)_{H}, \left(\half \right)_{V};   
\left(1\right)_{H} \left(1 \right)_{V} \rangle
$$
 The mutation matrix eq.(\ref{eq:mhv}) now becomes
 \begin{eqnarray*}
{M}&=&\sum_{m=H,V} \sum_{i=2,3} \mu_{1,m}[(A_{i,m}J_{-,m}+ J_{+,m} 
A_{i,m}^T) \\
& + & (J_{-,m}  A_{i,m}^{T} + A_{i,m} J_{+,m}) ]  
  +  \mu_{2,m} \, (J_{-,m} +   J_{+,m}) \\
  & + & \mu_{3,m }(  B_{m} J_{-,m} +  
 J_{+,m} B_{m}^T) + M_{D,m}
\end{eqnarray*} 
where
 \bea
 B_{m} \mid \mathbf{J} \rangle & = & \mid J_{3,m}, J^3_m, J^2_m  - 1 \rangle  \\
 A_{i,m} \mid \mathbf{J} \rangle &  = & \mid J_{3,m} ,J^{3}_m-1, \ldots 
J^{i}_m-1, \ldots \rangle
    \quad  (2 \leq i \leq 3)
 \eea
and $M_{D}$ is the diagonal part of the mutation matrix.
We are interested in finding the stationary  configuration solution of the eq.(\ref{eq:ME}) for the 
$64$ different possible sequences.  We choose
 the following form for the (purely additive) fitness
$H = J_{3,H} +  J_{3,V}  +  \lambda\mathbf{1}$, 
    $\lambda > 0$  ensuring    $H + M$ to be positive.
 Below we report several representative figures in which the obtained  numerical solutions  are fitted with a  function given by eq.(\ref{eq:bf}) (we omit the hat on the parameters).   In figg.\ref{Par96}-\ref{Par99}, with a suitable choice of 
the values of the parameters, our results are well fitted.
   In fig.\ref{Par94} we report another  solution where the ratio, denoted by $(H/V)$, between the mutation intensity between transitions and transversions, is chosen larger than one, but the value of the coupling constants do not satisfy the hierarchy $\mu_{1,H} \,  <  \, \mu_{2,H},  \mu_{3,H}$, which is less well fitted.   
 In fig.\ref{Par98} we report another  solution with a unrealistic choice of 
the values of the parameters of the ratio $(H/V)$ ($(H/V) \approx 10$), which is, indeed, badly fitted by a function given  by eq.(\ref{eq:bf}). Finally in fig.\ref{Par87} we report another solution, also badly fitted, where $(H/V) \approx 10^{-1}$.  This last result is a consequence of the fact that we have chosen a fitness symmetric for the exchange $H \llra V$. Therefore the exchange of the values of the coupling constants between $M_H$ and  $M_V$ gives the same shape of the distribution. Of course the rank of the same codon is, in general, different in the two cases. Summarizing, we can state that the  numerical solutions of our model, for arbitrary choice of the values of the coupling constants, are  rather well fitted by  a function of the type given in eq.(\ref{eq:bf}), with a suitabe choice of the parameters, but that
a non realistic choice  of the values of the coupling constants,  e.g. 
a ratio of transversion/transition
  mutation very high or very low, seems to destroy the goodness of the fit. Moreover,  it is  quite surprising to remark that the values of the parameters in the function  eq.(\ref{eq:bf}), which fits our numerical solutions, are of the same order of magnitude of the parameters (depending on the total $GC$ content) found in (Frappat, Minichini, Sciarrino and Sorba, 2003) to best fit the observed rank ordered distribution. In the present paper,  the values of $ \widehat{\alpha}$ and  $ \widehat{\eta}$   are found to be slightly larger   than the  ones computed  in (Frappat, Minichini, Sciarrino and Sorba, 2003). Let us stress once more that
  a mutation matrix  $M$, with non diagonal non vanishing entries 
connecting only codons with Hamming distance equal to one,   is unable to reproduce the observed rank ordered distribution as it induces mutation between  classes of codons  at the same Hamming  distance.
 We have considered separately the finess and mutation matrix for the {\it horizontal} and
   {\it vertical} labels of the codons.  As, a priori, one can consider also a coupling term between the two parts, our  simplified treatment has to be considered as a first step in the way of constructing a realistic model. We have  also performed a preliminary analysis with a a value, $\rho_H$, of the {\it horizontal}  fitness different from the value, $\rho_V$ of the  {\it vertical}  one. It appears that the  outcome depends on the ratio  $\rho_H/\rho_V$ as well as on the ratio between the values of the $\rho$ and the value of the transition coupling constant $\mu$ (beside the discussed dependence on the ratio of $\mu_H/\mu_V$ and on the hierarchy of the values of the differents $\mu_H$ and $\mu_V$). So we believe that 
  a better understanding of the form of the fitness and of the hierarchy of the values of the mutation parameters, as well as on the reliability of a model which explains the rank ordered distribution of the codons as a consequence of the mutation-selection of 64 triplets,  is necessary before  further pursuing the numerical analysis . 
  
\section{Conclusions}
 
\begin{figure}[tbh]
\begin{center}
\framebox{\epsfxsize=0.4\textwidth
\epsffile{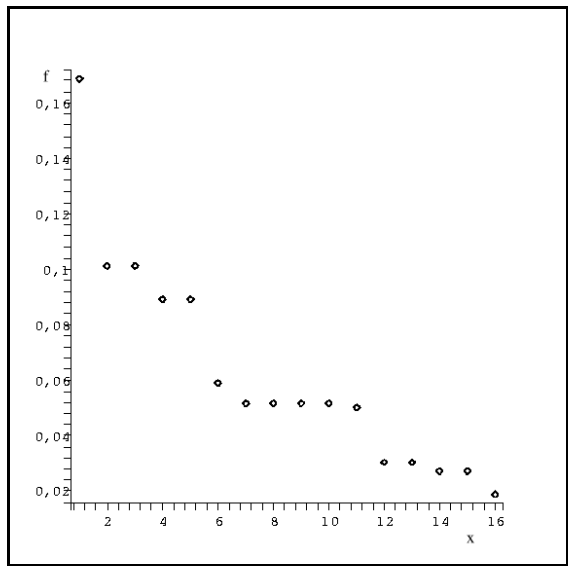}}
\end{center}
\caption{Rank ordered distribution of equilibrium population (N=4) obtained for a
transition matrix allowing the 
 same number of mutations as $M$, between sequences at different Hamming distances} 
\label{fig:cdc}
 \end{figure}

 We have proposed a model not analytically soluble, but  which admits an 
 easy numerical solution for short spin chains.
 Let us emphasize that the main purpose of the proposed scheme is to 
take into account, at least partially, the effects of the neighbours 
in the mutation.
 We  point out, once more, that our model is not equivalent to a model where
  the intensity  depends on the site  undergoing the 
  transition, or on the nature of the closest neighbours or on the 
  number of the $R$ and $Y$ labels of the sequence; indeed essentially 
  the intensity depends on the distribution in the sequence of 
  the $R$ and $Y$.
 We  find that the numerically computed stationary distribution  for short oligonucleotides to follow a Yule o Zipf law, in agreement with the observed distribution.
 We are far from claiming, for several obious reasons, that our simple  
 model is the only model able to explain the observed oligonucleotide distribution, but that the standard approach using the 
Hamming distance does not provide such a solution. 
One may correctly argue that the comparison between the Hamming model, 
depending on only one parameter and taking into account only one 
site spin flip, with our model, which depends on four parameters and 
takes into account spin flip of more than one site, is not meaningful. So we 
have computed the stationary distribution with a mutation matrix 
 not vanishing for Hamming distance larger than one and allowing the 
 same number of mutations as our model. The 
result reported in fig.\ref{fig:cdc} shows that the plateaux structure is 
always the dominant feature. Let us comment on the non point mutations which 
naturally are present in our model. In literature there is 
an increasing number of papers that, on the basis of more accurate data, 
question both the assumptions that mutations occur as single nucleotide
and as independent point event. In a quite recent paper 
(Whelan and Goldman, 2004)
  have presented a model allowing for single-nucleotide, 
doublet and triplet mutation, finding that the model provides 
statistically significant improvements in fits with protein coding 
sequences. We note that the triplet mutations, for which there is no known
inducing mechanism, but which can possibly be explained  by large 
scale event,  called sequence inversion
in (Whelan and Goldman, 2004), are indeed the kind of mutations, above discussed, that 
our model naturally describes.
 Doublet mutations do not appear, due to the assumed spin flip equal $\pm 1$, but
 on 
one side some of these mutations are hidden by the binary 
approximation, and on the other side the parameter  ruling such 
mutations, as computed in (Whelan and Goldman, 2004), is lower than the one ruling the triplet mutation.
 In conclusion the Hamming distance does not seem
 a suitable measure of the distance in the space of the biological sequences,
 the crystal basis, on the contrary, seems  a better candidate to parametrize
the elements of such space.
  Our model makes use of this parametrisation, allows to modelise
 some non point mutations and exhibits intriguing  and interesting features, 
hinting in the right direction, worthwhile to be further 
investigated.
In the present simple  version, the model depends 
only on 4 (8) parameters  in the two letter (resp. four letter) alphabet for any N, which are, very likely, not enough to 
describe sequences longer that the considered ones. However the model is rather
flexible: as shown in the case of the codons,  it is easily generalised to the four letter 
alphabet;  besides the obvious introduction of 
more coupling constants, it allows, e.g., to analyse part of the sequences containing 
hot spots in the mutation, to 
take into account doublet mutations (indeed the 
operator eq.(\ref{eq:ik}) or $A_{i,i+1}^{T}$ describes a doublet spin 
flip at position i,i+1).
Although the very short chain, which we were interested in, can be studied numerically 
without any use of the crystal basis,   we propose a 
general algorithm, which can be applied to chains of arbitrary 
length and which can be easily implemented in computers. 
   It is worthwhile to remark that we are trying to compare theoretical 
results, deriving from simple models, to really observed data, coming 
from the extremely complex biological world. In this context the 
crystal basis provides a compact and useful notation to describe the
``kinematical" variables which are changed by the dynamics.
  The generalisation of our approach to the a four letters alphabet, which is easily done replacing ${\cal U}_{q}(sl(2))$ 
by ${\cal U}_{q}(sl(2) \oplus sl(2))$,   has been presented and applied  to the study of the mutations of  the codons.
As expected, calculations  are more complicated and only a few results in the simple case of the triplets are given.  In this framework,   further investigation deserve attention, in particular to study oligonucleotide distribution in the four letters alphabet and mutations in long sequences.
In conclusion we point out that:
  \begin{itemize}
  \item the crystal basis  provides an alternative way of labelling 
 nucleotide sequences, in particular codons or genes, mapping any  finite ordered nucleotide sequence in a vector state of an irrep.. We  point out that  the choice of the
limit $q \to 0$  ({\bf crystal basis})  is  essential for the above identification as, only in this limit,
 due to Kashiwara theorem  (Kashiwara, 1990),
the composite states are pure states.
\item the mutation matrix  $M$ in our model
  describes an interaction on the $i$-th nucleotide depending on the 
input-output sequences and, in the flip of one spin (or double spin), inherently takes into account
 non local effects. 
 So the crystal basis variables are suitable to
partially describe non local events which affect the mutations. 
 \item models based on the crystal basis
seem, in the light of the obtained results,  better candidates than models based on Hamming distance to 
describe mutations. 

\end{itemize}
 
 As final remark, this article should be seen as a first, simplified attempt to build models, more realistic than the ones based on the Hamming distance, to describe the effects of the mutation-selection on the observed distribution of oligonucleotides.

\appendix
 
\section{Appendix A}

{\bf Label of N binary string}. Let us recall
 that the algebra 
${\cal U}_{q}(sl(2))$ is defined as a suitable completion of the algebra of 
polynomes in the generators $\widetilde{J}_{+}$, $\widetilde{J}_{-}$ and 
$\widetilde{J}_{3}$ (in particular adding the exponential series), subject 
to the following commutation relations:
\begin{eqnarray}
{}[\widetilde{J}_{+} ,\, \widetilde{J}_{-}] &=& [2 \widetilde{J}_{3}]_{q} 
\nonumber \\
{}[\widetilde{J}_{3} ,\, \widetilde{J}_{\pm}] &=& \pm \, \widetilde{J}_{\pm}
\end{eqnarray}
where 
\begin{equation}
{}[x]_q = \frac{q^x - q^{-x}}{q - q^{-1}} 
\end{equation}
Moreover some  more axioms have to be fulfilled, which endows ${\cal 
U}_{q}(sl(2))$ with a Hopf algebra structure. 
The vector spaces of the irreducible representations of this algebra are 
labelled, for $q$ different of root of unity, by a non negative integer or 
half-integer number $j$ and are of dimension $(2j+1)$, the basis vectors 
being denoted by $\psi_{jm}$, $-j \le m \le j$. In the limit $q \to 1$ one 
recovers the usual $sl(2)$. Strictly speaking, in the limit $q \to 0$ the 
generators are ill defined, but it is possible, see  (Kashiwara, 1990), to define 
new generators $J_{\pm}$, $J_{3} (= \widetilde{J}_{3})$, whose action on 
the vector basis of the representation space, still labelled by a non 
negative integer or half-integer number $j$ and of dimension $(2j + 1)$, is 
well defined:
\begin{equation}
\label{eq:action}
J_{3} \, \psi_{jm} = m \, \psi_{jm} \, \qquad J_{\pm} \, \psi_{jm} = 
\psi_{j,m \pm 1} \,  \qquad J_{\pm} \, \psi_{j, \pm j} = 0
\end{equation}
This special basis in the limit $q \to 0$ is called a crystal base. Note 
that the action of $J_{\pm}$ on $\psi_{jm}$ is equal to $\psi_{j,m \pm 1}$ 
(i.e. the coefficient is always 1), contrary to the $sl(2)$ or ${\cal 
U}_{q}(sl(2))$ case where this coefficient is a complicated function of $j$ 
and $m$. \\
It is possible also to define an operator $C$ called Casimir operator 
(Frappat, Sciarrino and Sorba, 1998) such that:
\begin{equation}
\label{eq:casimir}
C \, \psi_{jm} = j(j + 1) \, \psi_{jm} \;\; \Longrightarrow \;\; [C , \, 
J_{\pm}] = [C , \, J_{3}] = 0
\end{equation}
Its explicit expression is given by
\begin{equation}
\label{eq:casimir2}
C = (J_{3})^{2} + \half \sum_{n \in \ZZ_+} \sum_{k=0}^n (J_{-})^{n-k} 
(J_{+})^n (J_{-})^k 
\end{equation}
  In (Kashiwara, 1990) it has been shown
that the tensor product of two crystal bases labelled by $j_{1}$ and 
$j_{2}$ can be decomposed into a direct sum of crystal bases labelled, as 
in the case of the tensor product of two $sl(2)$ or of ${\cal U}_{q}(sl(2))$ irreducible representations, by an integer or half-integer 
number $j$ such that
\begin{equation}
\vert j_{1} - j_{2} \vert \le j \le j_{1} + j_{2}
\end{equation}
The new peculiar and crucial feature is that now the basis vectors of the
 $j$-space are \emph{pure states}, that is they are the product of a state belonging to 
the $j_{1}$-space and of a state belonging to the $j_{2}$-space, while in 
the case of $sl(2)$ or of ${\cal U}_{q}(sl(2))$ they are linear 
combinations with coefficients called respectively Clebsch-Gordan 
coefficients or $q$-Clebsch-Gordan coefficients.  Making use of this 
property any string of $N$ binary label (spin)  $ \; x \in \{\pm = 
R, Y \} \; $ can be seen as a state of an irreducible representation 
(irrep.) contained in $N$-fold tensor product of the the 2-dim fundamental 
irrep. (labelled by  $j = 1/2$) of $\; sl_{q  \to 0}(2) \;$  whose state are 
labelled by $ \; j_{3} =  \pm 1/2 = \pm = C, U \; $. Therefore, in the 
most general case, it can be identified by the following $N$ labels 
\begin{enumerate}
\item the value $\; J^{N} \; $ labelling the irrep. which the state belongs to.
This value is computed taking away  the  $Y$ elements, which are at right of $R$,
 contracting  each of them with  a $R$ element on the left, and then summing the numbers of left 
  $R$  and of  left  $Y$,  which are, respectively,  at  the right of $Y$ and at the left of $R$.
In other words this value is computed deleting all ordered couples $RY$ (of first neighbours) in the
sequence and iterating this procedure, on the generated sequence, up to no $Y$ are on the right of 
 any $R$. We refer to the elements which are deleted in the procedure as \emph{contracted}.
 
 \item the value of $J_{3}$, with $2J_{3} = n_{C} - n_{U}$, $n_{x}$ being the number of $x$ elements in the strings
 
\item the $N-2$ labels $J^{i}$ ($ 2 \leq i \leq N-1$), respectively identifying the irrep.,
which the sequence truncated to the $i$-th element belongs to.  
 \end{enumerate}
 E.g. the $N = 5$ string $CCUCU$ is labelled by: $ J^{5} = 1/2, J_{3} = 
 1/2,J^{4} = 1, J^{3} = 1/2, J^{2} = 1$; the string $CCCCC$ is labelled 
 by: $ J^{5} = 5/2, J_{3} = 5/2,J^{4} = 1, J^{3} = 1/2, J^{2} = 1$;
 the string $CUUCC$ is labelled 
 by: $ J^{5} = 3/2, J_{3} = 1/2, J^{4} = 1, J^{3} = 1/2, J^{2} = 0$.

{\bf Multiplicity of $sl(2)$ irreps. in the tensor product}.   It  is useful to know the number (multiplicity) 
 of irreps. 
labelled by the same value of $J$ appearing in the $N$-fold tensor 
product of the fundamental representanio $j = 1/2$, i.e. the 
coefficients $m_{i}$ appearing in the identity
\be
{\Large 
\otimes ^{N} \; {\bf 1/2}  = \otimes ^{N} \; J^{(1)} = J^{(N)}  \oplus 
\sum_{k \geq 1} \; m_{N - 2k}^{N} \; J^{(N - 2k)}}  \label{eq:t}
\ee
The number  $m_{N - 2k}^{N}$, giving the multiplicity of the irrep. 
$J^{(i)}$ in the tensor product  is given by   (Kirillov, 1991)
 \be
m_{N - 2k}^{N}  = \left [ \left( \begin{array}{c}
   m  \\  k    \end{array}\right)   \, - \, \left( \begin{array}{c}
   m  \\  k - 1     \end{array}\right)  \right]
\ee 
One gets:  
 \bea
&  m_{1}^{N} = 1 \nonumber \\ 
& m_{2}^{N} = N - 1 \nonumber \\
& m_{3}^{N} = \sum_{k = 2}^{N-2} \; k \;\;\;\;\;\; (N - 2 > 2) \nonumber \\
& m_{3 + j}^{N} = \sum_{K = 3+2j}^{N-1} \; m_{3 + j - 1}^{K} 
\;\;\;\;\;\; j > 0, \;\; N - 1 \geq 3 + 2j  \label{eq:mul}
\eea 
  E.g. for $ N =  6, 7, 8, 9$ we have, omitting the upper label 
 \bea
 &  m_{3} = 9 \;\;\;\;\;\; m_{4} = 5   \nonumber \\
&  m_{3} = 14 \;\;\;\;\;\; m_{4} = 14  \nonumber \\
&  m_{3} = 20 \;\;\;\;\;\; m_{4} = 28 \;\;\;\;\;\; m_{5} = 14 
  \nonumber \\
& m_{3} = 27 \;\;\;\;\;\; m_{4} = 48 \;\;\;\;\;\; m_{5} = 42
 \eea
(Recall that  $ J^{(i)} = \bf{i/2}$)

 
 \section*{References}

\begin{trivlist}
 
 \item 
   Arndt P.F., Burge C.B. and Hwa  T., 2002,  \textsl{DNA Sequence Evolution with Neighbor-Dependent  Mutation},  RECOMB2002, Proc. 6th Int.Conf. on Computational Biology (2002), p.32 (physics/0112029).
 
 \item 
  Blake R.D., Hess S.T. and  Nicholson-Tuell J., 1992,  \textsl{ The influence of Nearest Neighbors on the Rate and Pattern of Spontaneous Mutations}, J.Mol.Evol. \textbf{34}, 189.

 \item
  Blake R.D., Hess S.T. and  Blake R.D.,   1994, \textsl{ Wide Variations in Neighbor-dependent Substitution Rates }, J.Mol.Biol. \textbf{236}, 1022.
 
 \item 
   Baake  E.,  Baake M. and  Wagner H., 1997, \textit{Ising quantum  chain is equivalent to a model of biologicalevolution}, Phys.Rev.Lett. \textbf{78} , 559; Erratum Phys.Rev.Lett. \textbf{79} , 1782.
 
  \item
   Baake  E.,  Baake M. and  Wagner H., 1998,   \textit{Quantum mechanics versus classical probability in biological evolution}, Phys.Rev. E \textbf{57}, 1191. 
   
    \item 
 Baake  E. and Wagner H., 2001, \textit{Mutation-selection models solved exactly with methods of statistical mechanics},  Genet.Res.Camb. \textbf{78}, 93.
    
    \item 
 Blake  R.D., Hess S.T.  and  Nicholson-Tuell J., 1992, \textit{ The Influence of Nearest Neighbors on the Rate and Pattern of Spontaneous Point Mutation}, J.Mol.Evol. \textbf{34}, 189.
  
 \item
       Eigen M., 1971,  \textit{Selforganisation of matter and the evolution of biological macromolecules}, Naturwissenschaften \textbf{58},  465.  
  
  \item 
       Eigen M.,  McCaskill  J. and Schuster P., 1989, \textit{The molecular quasi-species},  J.Chem.Phys. \textbf{75},  149.
 
  \item
 \textit{Encyclopedic Dictionary of Mathematics}, 1960, 2nd edition, The MIT Press, Cambridge, (MA).
 
 \item
 Frappat L., Sciarrino  A. and Sorba P., 1998, \textsl{ A crystal basis model of the 
genetic code}  Phys.  Lett. A  \textbf{250}  214.  
 
 \item
 Frappat L., Sciarrino  A. and Sorba P., 2001, \textsl{ Crystalizing  the 
genetic code} J.Biol. Phys.   \textbf{27}  1.  
 
  \item
   Frappat L.,  Minichini C., Sciarrino  A. and Sorba P., 2003,
     \textit{Universality and Shannon Entropy for Codon Usage} 
 Phys.Rev. E    \textbf{68} , 061910.  

\item
    Goldstein RF. and Benight A. S., 1992, \textit{How Many Numbers Are Required to Specify Sequence-Dependent Properties of Polynucleotides?}, Biopolymers \textbf{32}, 1679.
    
 \item 
   Hermisson J., Wagner  H. and Baake M., 2001, \textit{Four-State Quantum Chain as a Model of Sequence Evolution},  J.Stat.Phys.  \textbf{102}, 315. 
 
  \item 
  Hermisson J., Redner O. , Wagner  H. and Baake E., 2002,  textit{Mutation-selection balance: Ancestry, load and maximum priciple}, Theoretical Population Biology  \textbf{62}, 9. 
  
  \item 
 Hess S.T., Blake  J.D. and  Blake  R.D., 1994, \textit{Wide Variations in Neighbor-dependent Substitution  Rates}, J.Mol.Evol. \textbf{236}, 1022.
  
    \item 
    Hofbauer  J. and Sigmund K., 1988, {\em The Theory of Evolution and Dynamical
    Systems}. Cambridge University Press, Cambridge.

    \item
Kashiwara M., 1990, \textsl{Crystallizing the $q$-analogue of universal 
enveloping algebras,}  Commun.  Math.  Phys.  \textbf{133} 249.

\item 
 Kirillov A.A., 1991, {\em Representation theory and Noncommutative 
Harmonic Analysisis},  Encyclopedia of Mathematical 
Sciences, Vol. 22, pag. 70, Springer Verla, Berlin.

   \item 
    Leuth\"{a}usser I., 1986, \textit{An exact correspondence between Eigen's evolution model and a two-dimensional Ising system},  J.Chem.Phys. \textbf{84}, 1884.  
 
  \item 
    Leuth$\ddot{a}$usser I., 1987, \textit{Statistical mechanics of Eigen's evolution model},   J.Stat.Phys. \textbf{48}, 343. 

  \item
   Li Wen-Hsiung, 1997, {\em Molecular Evolution}, Sinauer Associates 
Incorporated, Sunderland.
 
  \item 
Martindale C. and Konopka A. K.,  1996, \textit{Oligonucleotide Frequencies in DNA Follow a Yule Distributiojn}, Computers Chem. \textbf{20}, 35.
  
 \item
   Minichini C.  and Sciarrino  A., 2004a, 
   \textsl{ Quantum Spin Model fitting the Yule distribution of oligonucleotides },
 quant-ph/0409071
 
 \item
   Minichini C.  and Sciarrino  A., 2004b, 
   \textsl{ Mutation Model fitting the Yule distribution of oligonucleotides}, q-bio.BM/0412006
    
       \item 
  Saakian S. and Hu Chin-Kun, 2004, \textit{Eigen Model as a Quantum Spin Chain: Exact Dynamics}, Phys.Rev. E \textbf{69}, 021913 (cond-mat/0402212). 
 
  \item 
SantaLucia, J., 1998, \textit{A unified view of polymer, dumbbell, and oligonucleotide DNA nearest-neighbor thermodynamics},  Proc.Natl.Acad.Sci USA \textbf{95}, 1460.

  \item 
 Wagner  H., Baake E. and Gerisch T., 1998, \textit{Ising quantum chain and sequence evolution}, J.Stat.Phys. \textbf{92} , 1017. 
   
 \item
 Whelan S. and Goldman N., 2004, \textit{Estimating the Frequency of Events That Cause Multiple-Nucleotide Changes}, Genetics \textbf{167}, 2027.
 
   \item 
  Xia  Tiambing et al,  1998, \textit{Thermodynamic Parameters for an Expanded Nearest-Neighbor Model for Formation of RNA Duplexes with Watson-Crick Base Pairs}, Biochemistry \textbf{37}, 14719.  
     
    \item 
Yule G.U., 1924,  \textit{A mathematical theory of evolution,  based on the conclusions of Dr.J.C. Willis, F.R.S.}, Phil.Trans. B \textbf{213},  21.
      
     \item 
Zipf, G.K., 1949  \em{Human behaviour and the Principle of Least Effort}, Addison-Wesley Press, Cambridge, MA .

\end{trivlist}
   
\newpage
 
\end{document}